# DECAYS OF BOTTOM MESONS EMITTING PSEUDOSCALAR AND TENSOR MESONS IN ISGW II MODEL

Neelesh Sharma and R.C. Verma
Department of Physics, Punjabi University,
Patiala-147 002, INDIA.

**Abstract**

In this paper, we investigate phenomenologically two-body weak decays of the bottom mesons emitting pseudoscalar and tensor mesons. Decay amplitudes are obtained using the factorization scheme in the improved ISGW II model. Branching ratios for the CKM-favored and CKM-suppressed decays are calculated.

# I. INTRODUCTION

Experimental results are available for the branching ratios of several $B$-meson decay modes. Many theoretical works have been done to understand exclusive hadronic $B$ decays in the framework of the generalized factorization, QCD factorization or flavor SU(3) symmetry. Weak hadronic decays of the $B$-mesons are expected to provide a rich phenomenology yielding a wealth of information for testing the standard model and for probing strong interaction dynamics. However, these decays involve nonperturbative strong processes which cannot be calculated from the first principles. Thus phenomenological approaches [1-5] have generally been applied to study them using factorization hypothesis. It involves the expansion of the transition amplitudes in terms of a few invariant form factors which provide essential information on the structure of the mesons and the interplay of the strong and weak interactions. This scheme has earlier been employed to study the weak hadronic decays of $B$-meson to $s$-wave mesons [5-12]. $B$-mesons, being heavy, can also emit heavier mesons such as $p$-wave mesons, which have attracted theoretical attention recently. However, there exist a few works on the hadronic $B$ decays [13-17] that involve a tensor meson in the final state using the frameworks of flavor SU(3) symmetry and the generalized factorization. In the next few years new experimental data on rare decays of $B$ mesons would become available from the $B$ factories such as Belle, Babar, BTeV, LHC. It is expected that improved measurements or new bounds will be obtained on the branching ratios for various decay modes and many decay modes with small branching ratios may also be observed for the first time.

In this paper, we analyze two-body hadronic decays of $B^-$, $\overline{B}^0$ and $\overline{B}_s^0$ mesons to pseudoscalar ($P$ ($0^-$)) meson and tensor ($T$ ($2^+$)) meson, for whom the experiments have provided the following branching ratios [18,19]:

$$B(B^- \to \pi^- D_2^0) = (7.8 \pm 1.4) \times 10^{-4},$$
$$B(B^- \to \pi^- f_2) = (8.2 \pm 2.5) \times 10^{-6},$$
$$B(B^- \to K^- f_2) = (1.3^{+0.4}_{-0.5}) \times 10^{-6},$$
$$B(B^- \to \eta K_2^-) = (9.1 \pm 3.0) \times 10^{-6},$$
$$B(\overline{B}^0 \to \eta \overline{K}_2^0) = (9.6 \pm 2.1) \times 10^{-6},$$
$$B(\overline{B}^0 \to \overline{D}^0 f_2) = (1.2 \pm 0.4) \times 10^{-4},$$
$$B(\overline{B}^0 \to \pi^{\mp} a_2^{\pm}) = < 3.0 \times 10^{-4}, \qquad (1)$$
$$B(B^- \to \pi^- \overline{K}_2^0) = < 6.9 \times 10^{-6},$$
$$B(\overline{B}^0 \to D_s^- a_2^+) = < 1.9 \times 10^{-4},$$
$$B(\overline{B}^0 \to \pi^+ K_2^-) = < 1.8 \times 10^{-5},$$
$$B(\overline{B}^0 \to \pi^- D_2^+) = < 2.2 \times 10^{-3}.$$

Employing the factorization scheme, we calculate the decay amplitudes for CKM-favored and CKM-suppressed modes involving $b \to c$ and $b \to u$ transitions in the Isgur-Scora-Grinstein-Wise (ISGW II) model [2, 3]. In general, $W$-annihilation and $W$-exchange diagrams may also contribute to these decays under consideration. Normally, such contributions are expected to be suppressed due to the helicity and color arguments and are neglected in this work.

The paper is organized as follows: In Sec. II, we present meson spectroscopy. Methodology for calculating $B \to PT$ is provided in Sec. III. Sec. IV deals with numerical results and discussions. Summary and conclusions are given in the last section.

## II. MESON SPECTROSCOPY

Experimentally [18], the tensor meson sixteen-plet comprises of an isovector $a_2(1.318)$, strange isospinor $K_2^*(1.429)$, charm SU(3) triplet $D_2^*(2.457)$, $D_{s2}^*(2.573)$ and three isoscalars $f_2(1.275)$, $f_2'(1.525)$ and $\chi_{c2}(3.555)$. These states behave well with respect to the quark model assignments, though the spin and parity of the charm isosinglet $D_{s2}^*(2.573)$ remain to be confirmed. The numbers given within parentheses indicate mass (in GeV units) of the respective mesons. $\chi_{c2}(3.555)$ is assumed to be pure $(c\bar{c})$ state, and mixing of the isoscalar states is defined as:

$$f_2(1.275) = \frac{1}{\sqrt{2}}(u\bar{u} + d\bar{d})\cos\phi_T + (s\bar{s})\sin\phi_T,$$
$$f_2'(1.525) \frac{1}{\sqrt{2}}(u\bar{u} + d\bar{d})\sin\phi_T - (s\bar{s})\cos\phi_T, \qquad (2)$$

where $\phi_T = \theta(ideal) - \theta_T(physical)$ and $\theta_T(physical) = 27°$ [18].

Similarly, for $\eta$ and $\eta'$ states of well established pseudoscalar sixteen-plet, we use

$$\eta(0.547) = \frac{1}{\sqrt{2}}(u\bar{u} + d\bar{d})\sin\phi_P - (s\bar{s})\cos\phi_P,$$
$$\eta'(0.958) = \frac{1}{\sqrt{2}}(u\bar{u} + d\bar{d})\cos\phi_P + (s\bar{s})\sin\phi_P, \qquad (3)$$

where $\phi_p = \theta(ideal) - \theta_p(physical)$ and we take $\theta_p(physical) = -15.4°$ [18]. $\eta_c$ is taken as

$$\eta_c(2.979) = (c\bar{c}). \qquad (4)$$

## III. METHODOLOGY

### A. Weak Hamiltonian

For bottom changing $\Delta b = 1$ decays, the weak Hamiltonian involves the bottom changing current,

$$J_\mu = (\bar{c}b)V_{cb} + (\bar{u}b)V_{ub}, \qquad (5)$$

where $(\bar{q}_i q_j) \equiv \bar{q}_i \gamma_\mu (1-\gamma_5) q_j$ denotes the weak V-A current. QCD modified weak Hamiltonian is then given below:

i) for decays involving $b \to c$ transition,

$$\begin{aligned}
H_W = \frac{G_F}{\sqrt{2}} \{ & V_{cb} V_{ud}^* [a_1 (\bar{c}b)(\bar{d}u) + a_2 (\bar{d}b)(\bar{c}u)] + \\
& V_{cb} V_{cs}^* [a_1 (\bar{c}b)(\bar{s}c) + a_2 (\bar{s}b)(\bar{c}c)] + \\
& V_{cb} V_{us}^* [a_1 (\bar{c}b)(\bar{s}u) + a_2 (\bar{s}b)(\bar{c}u)] + \\
& V_{cb} V_{cd}^* [a_1 (\bar{c}b)(\bar{d}c) + a_2 (\bar{d}b)(\bar{c}c)] \},
\end{aligned} \qquad (6a)$$

ii) for decays involving $b \to u$ transition,

$$\begin{aligned}
H_W = \frac{G_F}{\sqrt{2}} \{ & V_{ub} V_{cs}^* [a_1 (\bar{u}b)(\bar{s}c) + a_2 (\bar{s}b)(\bar{u}c)] + \\
& V_{ub} V_{ud}^* [a_1 (\bar{u}b)(\bar{d}u) + a_2 (\bar{d}b)(\bar{u}u)] + \\
& V_{ub} V_{us}^* [a_1 (\bar{u}b)(\bar{s}u) + a_2 (\bar{s}b)(\bar{u}u)] + \\
& V_{ub} V_{cd}^* [a_1 (\bar{u}b)(\bar{d}c) + a_2 (\bar{d}b)(\bar{u}c)] \},
\end{aligned} \qquad (6b)$$

where $\bar{q}_i q_j \equiv \bar{q}_i \gamma_\mu (1-\gamma_5) q_j$ denotes the weak V-A current and $V_{ij}$ are the well-known CKM matrix elements, $a_1$ and $a_2$ are the QCD coefficients. By factorizing matrix elements of the four-quark operator contained in the effective Hamiltonian (6), one can distinguish three classes of decays [20]:

- class I transition <u>caused by color favored diagram</u>: the corresponding decay amplitudes are proportional to $a_1$, where $a_1(\mu) = c_1(\mu) + \frac{1}{N_c} c_2(\mu)$, and $N_c$ is the number of colors.

- class II transition <u>caused by color suppressed diagram</u>: the corresponding decay amplitudes in this class are proportional to $a_2$ i.e. for the color suppressed modes

$$a_2(\mu) = c_2(\mu) + \frac{1}{N_c} c_1(\mu).$$

- class III transition <u>caused by both color favored and color suppressed diagrams</u>: these decays experience the interference of color favored and color suppressed diagrams.

We follow the general convention of large $N_c$ limit to fix the QCD coefficients $a_1 \approx c_1$ and $a_2 \approx c_2$, where [20,21]:

$$c_1(\mu) = 1.12 \ , \ c_2(\mu) = -0.26 \text{ at } \mu \approx m_b^2. \tag{7}$$

### B. Decay Amplitudes and Rates

The decay rate formula for $B \rightarrow PT$ decays is given by

$$\Gamma(B \rightarrow PT) = \left(\frac{m_B}{m_T}\right)^2 \frac{p_C^5}{12\pi m_T^2} |A(B \rightarrow PT)|^2, \tag{8}$$

where $p_C$ is the magnitude of the three-momentum of the final-state particle in the rest frame of $B$-meson and $m_B$ and $m_T$ denote masses of the $B$-meson and tensor meson, respectively.

The factorization scheme in general expresses the weak decay amplitude as the product of matrix elements of weak currents (up to the weak scale factor of $\frac{G_F}{\sqrt{2}} \times CKM$ elements $\times$ QCD factor),

$$\langle PT | H_W | B \rangle \approx \langle P | J^\mu | 0 \rangle \langle T | J_\mu | B \rangle + \langle T | J^\mu | 0 \rangle \langle P | J_\mu | B \rangle. \tag{9}$$

However, the matrix elements $\langle T(q_\mu) | J_\mu | 0 \rangle$ vanish due to the tracelessness of the polarization tensor $\epsilon_{\mu\upsilon}$ of spin 2 meson and the auxiliary condition $q^\mu \epsilon_{\mu\upsilon} = 0$ [19]. Remaining matrix elements are expressed as:

$$\langle P(k_\mu) | J_\mu | 0 \rangle = -if_P k_\mu,$$
$$\langle T(P_T) | J_\mu | B(P_B) \rangle = ih \epsilon_{\mu\upsilon\lambda\rho} \epsilon^{*\upsilon\alpha} P_{B\alpha} (P_B + P_T)^\lambda (P_B - P_T)^\rho + k \epsilon^*_{\mu\upsilon} P_B^\upsilon$$
$$+ b_+ (\epsilon^*_{\alpha\beta} P_B^\alpha P_B^\beta)[(P_B + P_T)_\mu + b_- (P_B - P_T)_\mu], \tag{10}$$

in the ISGW model [3] which yields

$$\langle PT|H_W|B\rangle = -if_P(\epsilon^*_{\mu\nu} P_B^\mu P_B^\nu) F^{B\to T}(m_P^2), \tag{11}$$

where

$$F^{B\to T}(m_P^2) = k(m_P^2) + (m_B^2 - m_T^2) b_+(m_P^2) + m_P^2 b_-(m_P^2). \tag{12}$$

Thus

$$A(B\to PT) = \frac{G_F}{\sqrt{2}} \times (CKM\ factors \times QCD\ factors \times CG\ factors) \times f_P F^{B\to T}(m_P^2). \tag{13}$$

### C. Form Factors in the ISGW II Model

The form factors have the following expressions in the ISGW II quark model, for $B\to T$ transitions [3]:

$$k = \frac{m_d}{\sqrt{2}\beta_B}(1+\tilde{\omega})\ F_5^{(k)},$$

$$b_+ + b_- = \frac{m_d}{4\sqrt{2}m_d m_b \tilde{m}_B \beta_B} \frac{\beta_T^2}{\beta_{BT}^2}\left(1 - \frac{m_d}{2\tilde{m}_B}\frac{\beta_T^2}{\beta_{BT}^2}\right) F_5^{(b_+ + b_-)},$$

$$\tag{14}$$

$$b_+ - b_- = -\frac{m_d}{\sqrt{2}m_b \tilde{m}_T \beta_B}\left(1 - \frac{m_d m_b}{2\mu_+ \tilde{m}_B}\frac{\beta_T^2}{\beta_{BT}^2} + \frac{\beta_T^2}{4\beta_{BT}^2}\left(1 - \frac{m_d}{2\tilde{m}_B}\frac{\beta_T^2}{\beta_{BT}^2}\right)\right) F_5^{(b_+ - b_-)},$$

where

$$F_5^{(k)} = F_5 \left(\frac{\overline{m}_B}{\tilde{m}_B}\right)^{-1/2} \left(\frac{\overline{m}_T}{\tilde{m}_T}\right)^{1/2},$$

$$F_5^{(b_+ + b_-)} = F_5 \left(\frac{\overline{m}_B}{\tilde{m}_B}\right)^{-5/2} \left(\frac{\overline{m}_T}{\tilde{m}_T}\right)^{1/2}, \tag{15}$$

$$F_5^{(b_+ - b_-)} = F_5 \left(\frac{\overline{m}_B}{\tilde{m}_B}\right)^{-3/2} \left(\frac{\overline{m}_T}{\tilde{m}_T}\right)^{-1/2},$$

$t(\equiv q^2)$ dependence is given by

$$\tilde{\omega} - 1 = \frac{t_m - t}{2\overline{m}_B \overline{m}_T}, \tag{16}$$

and the common scale factor

$$F_5 = \left(\frac{\tilde{m}_T}{\tilde{m}_B}\right)^{1/2} \left(\frac{\beta_T \beta_B}{\beta_{BT}^2}\right)^{5/2} \left[1 + \frac{1}{18} h^2 (t_m - t)\right]^{-3}, \qquad (17)$$

where

$$h^2 = \frac{3}{4 m_b m_q} + \frac{3 m_d^2}{2 \bar{m}_B \bar{m}_T \beta_{BT}^2} + \frac{1}{\bar{m}_B \bar{m}_T} \left(\frac{16}{33 - 2 n_f}\right) \ln\left[\frac{\alpha_S(\mu_{QM})}{\alpha_S(m_q)}\right], \qquad (18)$$

and

$$\beta_{BT}^2 = \frac{1}{2} \left(\beta_B^2 + \beta_T^2\right).$$

$\tilde{m}$ is the sum of the mesons constituent quarks masses, $\bar{m}$ is the hyperfine averaged physical masses, $n_f$ is the number of active flavors, which is taken to be five in the present case, $t_m = (m_B - m_T)^2$ is the maximum momentum transfer and

$$\mu_+ = \left(\frac{1}{m_d} + \frac{1}{m_b}\right)^{-1}, \qquad (19)$$

Here, $m_d$ is the spectator quark mass in the decaying particle. For $B_s \to T$ transitions, $m_d$ is replaced with $m_s$. We take the following constituent quark masses (in GeV):

$$m_u = m_d = 0.33, \ m_s = 0.55, \ m_c = 1.82, \ m_b = 5.20, \qquad (20)$$

which are taken from the ISGW II model [3] in which treats mesons as composed of the constituent quarks. Values of the parameter $\beta$ for different $s$-wave and $p$-wave mesons are given in the Table I. We obtain the form factors describing $B \to T$ transitions which are given in Table II at $q^2 = t_m$. For the sake of comparison, the form factors obtained in the ISGW I quark model [2], are given in parentheses in Table II.

## IV. NUMERICAL RESULTS AND DISCUSSIONS

Sandwiching the weak Hamiltonian (6) between the initial and final states, we obtain decay amplitudes of $B^-$, $\bar{B}^0$ and $\bar{B}_s^0$ mesons for various decay modes as given in the Tables III, IV, V(a) and V(b). For numerical calculations, we use the following values of the decay constants (given in GeV) of the pseudoscalar mesons [13, 18, 21]:

$$f_\pi = 0.131, \ f_K = 0.160, \ f_D = 0.223, \ f_{D_s} = 0.294,$$
$$f_\eta = 0.133, \ f_{\eta'} = 0.126 \text{ and } f_{\eta_c} = 0.400. \qquad (21)$$

Finally, we calculate branching ratios of B-meson decays in CKM-favored and CKM-suppressed modes involving $b \to c$ and $b \to u$ transitions. The results are given in column III of the Tables VI, VII, VIII(a) and VIII(b) for various possible modes. We make the following observations:

1. **$B \to PT$ decays involving $b \to c$ transition**

   a) $\Delta b = 1, \Delta C = 1, \Delta S = 0$ mode:

      i. Calculated branching ratio $B(B^- \to \pi^- D_2^0) = 6.7 \times 10^{-4}$ agrees well with the experiment value [19] $(7.8 \pm 1.4) \times 10^{-4}$, and $B(\bar{B}^0 \to \pi^- D_2^+) = 6.1 \times 10^{-4}$, is well below the experimental upper limit $< 2.2 \times 10^{-3}$.

      ii. Branching ratios of other dominant modes are $B(B^- \to D^0 a_2^-) = 1.8 \times 10^{-4}$, $B(\bar{B}_s^0 \to \pi^- D_{s2}^+) = 7.1 \times 10^{-4}$, and $B(\bar{B}_s^0 \to D^0 K_2^0) = 1.1 \times 10^{-4}$. We hope that these values are within the reach of the furure experiments.

      iii. Decays $\bar{B}^0 \to D^0 a_2^0$ and $\bar{B}^0 \to D^0 f_2$ have branching ratios of the order of $10^{-5}$, since these involve color-suppressed spectator process. The branching ratio of $\bar{B}^0 \to D^0 f_2'$ decay is further suppressed due to the $f_2 - f_2'$ mixing being close to the ideal mixing.

      iv. Decays $\bar{B}^0 \to \pi^0 D_2^0 / \eta D_2^0 / \eta' D_2^0 / D^+ a_2^- / D_s^+ K_2^- / K^- D_{s2}^+$ and $\bar{B}_s^0 \to K^0 D_2^0 / D_s^+ a_2^-$ are forbidden in the present analysis due to the vanishing matrix element between the vacuum and tensor meson. However, these may occur through an annihilation mechanism. The decays $\bar{B}^0 \to \pi^0 D_2^0 / D^+ a_2^-$ may also occur through elastic final state interactions (FSIs).

   b) $\Delta b = 1, \Delta C = 0, \Delta S = -1$ mode:

      i. Dominant modes are found to have branching ratios: $B(B^- \to D_s^- D_2^0) = 6.8 \times 10^{-4}$, $B(B^- \to \eta_c K_2^-) = 1.4 \times 10^{-4}$,

$$B(\bar{B}^0 \to D_s^- D_2^+) = 6.4 \times 10^{-4}, \quad B(\bar{B}^0 \to \eta_c \bar{K}_2^0) = 1.3 \times 10^{-4},$$
$$B(\bar{B}_s^0 \to D_s^- D_{s2}^-) = 7.7 \times 10^{-4} \text{ and } B(\bar{B}_s^0 \to \eta_c f_2') = 1.3 \times 10^{-4}.$$

ii. Decays $B^- \to D^0 D_{s2}^- / D_s^- D_2^0 / K^- \chi_{c2}(1P)$, $\bar{B}^0 \to D^+ D_{s2}^- / D_s^- D_2^+ / K^- \chi_{c2}(1P)$ and $\bar{B}_s^0 \to \pi^0 \chi_{c2} / \eta \chi_{c2} / \eta' \chi_{c2} / D^+ D_2^- / D^0 \bar{D}_2^0 / D_s^+ D_{s2}^- / D^- D_2^+ / \bar{D}^0 D_2^0 / \eta_c a_2^0$ are forbidden in our work. Penguin diagrams may cause $B^- \to D^0 D_{s2}^- / D_s^- D_2^0$ and $\bar{B}^0 \to D^+ D_{s2}^- / D_s^- D_2^+$ decays, however these are likely to remain suppressed as these decays require $c\bar{c}$ pair to be created.

c) $\Delta b = 1, \Delta C = 0, \Delta S = 0$ mode:

i. For dominant decays, we predict $B(B^- \to D^- D_2^0) = 2.5 \times 10^{-5}$, $B(\bar{B}^0 \to D^- D_2^+) = 2.4 \times 10^{-5}$ and $B(\bar{B}_s^0 \to D^- D_{s2}^+) = 2.9 \times 10^{-5}$.

ii. Decays $B^- \to D^0 D_2^- / \pi^- \chi_{c2}(1P)$, $\bar{B}^0 \to D^0 \bar{D}_2^0 / D_s^- D_{s2}^+ / D^+ D_2^- / \bar{D}^0 D_2^0 / D_s^+ D_{s2}^- / \pi^0 \chi_{c2}(1P) / \eta \chi_{c2}(1P) / \eta' \chi_{c2}(1P)$ and $\bar{B}_s^0 \to K^0 \chi_{c2} / D_s^+ D_2^-$ are forbidden in our analysis. Annihilation diagrams, elastic FSI and penguin diagrams may generate these decays to the naked charm mesons. However, decays emitting charmonium $\chi_{c2}(1P)$ remains forbidden in the ideal mixing limit.

d) $\Delta b = 1, \Delta C = 1, \Delta S = -1$ mode:

i. Branching ratios of the dominant decays are $B(B^- \to K^- D_2^0) = 4.8 \times 10^{-5}$, $B(\bar{B}^0 \to K^- D_2^+) = 4.5 \times 10^{-5}$ and $B(\bar{B}_s^0 \to K^- D_{s2}^+) = 5.2 \times 10^{-5}$.

ii. Decays $\bar{B}^0 \to \bar{K}^0 D_2^0 / D^+ K_2^-$ and $\bar{B}_s^0 \to \pi^- D_2^+ / \pi^0 D_2^0 / \eta D_2^0 / \eta' D_2^0 / D^+ a_2^- / D^0 a_2^0 / D_s^+ K_2^-$ are forbidden in our analysis. Annihilation diagrams do not contribute to these decays. However, these may acquire nonzero branching ratios through elastic FSI.

## 2. $B \to PT$ decays involving $b \to u$ transition

a) $\Delta b = 1, \Delta C = 0, \Delta S = 0$ mode:

  i. $B(B^- \to \pi^- f_2) = 7.1 \times 10^{-6}$ is in good agreement with the experimental value $(8.2 \pm 2.5) \times 10^{-6}$ and $B(\bar{B}^0 \to \pi^- a_2^+) = 1.3 \times 10^{-5}$ is well below the experimental upper limit $< 3.0 \times 10^{-4}$.

  ii. $B^- \to K^0 K_2^- / K^- K_2^0$, $\bar{B}^0 \to K^+ K_2^- / K^0 \bar{K}_2^0 / K^0 \bar{K}_2^0 / \bar{K}^0 K_2^0 / K^- K_2^+ / \pi^+ a_2^-$ and $\bar{B}_s^0 \to K^+ a_2^- / K^0 a_2^0 / K^0 f_2 / K^0 f_2'$ are forbidden in the present analysis. Annihilation process and FSIs may generate these decays.

  iii. $B^- \to \pi^- \bar{K}_2^0$ and $\bar{B}^0 \to \pi^+ K_2^-$ are also forbidden in the present analysis which may be generated through annihilation diagram or elastic FSI.

b) $\Delta b = 1, \Delta C = -1, \Delta S = -1$ mode:

  i. Branching ratios $B(B^- \to D_s^- a_2^0) = 2.0 \times 10^{-5}$, $B(B^- \to D_s^- f_2') = 2.2 \times 10^{-5}$, $B(\bar{B}^0 \to D_s^- a_2^+) = 3.8 \times 10^{-5}$ and $B(\bar{B}_s^0 \to D_s^- K_2^+) = 2.6 \times 10^{-5}$ have relatively large branching ratios.

  ii. Decays $B^- \to \pi^0 D_{s2}^- / \eta D_{s2}^- / \eta' D_{s2}^- / \bar{K}^0 D_2^- / K^- D_2^0 / D^- \bar{K}_2^0$, $\bar{B}^0 \to \bar{K}^0 D_2^0 / \pi^+ D_{s2}^-$ and $\bar{B}_s^0 \to K^+ D_{s2}^- / \pi^+ D_2^- / \pi^0 \bar{D}_2^0 / \eta \bar{D}_2^0 / \eta' \bar{D}_2^0 / \bar{D}^0 a_2^0 / D^- a_2^+$ are forbidden in the present analysis. Annihilation and FSIs may generate these decays.

c) $\Delta b = 1, \Delta C = -1, \Delta S = 0$ mode:

  i. Branching ratios of $B(\bar{B}^0 \to \bar{D}^0 f_2) = 3.6 \times 10^{-8}$ is smaller than the experimental value $(1.2 \pm 0.4) \times 10^{-4}$. It may be noted that W-annihilation and W-exchange diagrams may also contribute to the B decays under consideration. Normally, such contributions are expected to be suppressed due to the helicity and color arguments. Including the factorizable contribution of such diagrams, the decay amplitude of $\bar{B}^0 \to \bar{D}^0 f_2$ get modified to (leaving aside the scale factor $\frac{G_F}{\sqrt{2}} V_{ub} V_{cd}^*$)

$$A(\bar{B}^0 \to \bar{D}^0 f_2) = \frac{1}{\sqrt{2}} a_2 f_D \cos\phi_T F^{B \to f_2}(m_D^2) +$$
$$\frac{1}{\sqrt{2}} a_2 f_B \cos\phi_T F^{f_2 \to D}(m_B^2). \quad (22)$$

Using $f_B = 0.176$ GeV, we find that the experimental branching ratio $B(\bar{B}^0 \to \bar{D}^0 f_2)$ requires $F^{f_2 \to D}(m_B^2) = -9.99$ GeV. This in turn enhances the branching ratio for $B^- \to D^- f_2$ to $1.2 \times 10^{-4}$.

ii. Dominant decay is $B(\bar{B}^0 \to D^- a_2^+) = 1.2 \times 10^{-6}$ and next order dominant decays are $B(B^- \to D^- f_2) = 6.9 \times 10^{-7}$ $B(B^- \to D^- a_2^0) = 6.5 \times 10^{-7}$ and $B(\bar{B}_s^0 \to D^- K_2^{*+}) = 8.3 \times 10^{-7}$.

iii. Decays $B^- \to K^0 D_{s2}^- / \pi^0 D_2^- / \pi^- \bar{D}_2^0 / \eta D_2^- / \eta' D_2^- / D_s^- K_2^0 / \eta_c D_2^-$, $\bar{B}^0 \to K^+ D_{s2}^- / \pi^+ D_2^- / \pi^0 \bar{D}_2^0 / \eta \bar{D}_2^0 / \eta' \bar{D}_2^0 / D_s^- K_2^+ / \eta_c D_2^0$ and $\bar{B}_s^0 \to K^0 \bar{D}_2^0$ are forbidden in the present analysis. Annihilation diagrams may generate these decays.

d) $\underline{\Delta b = 1, \Delta C = 0, \Delta S = -1 \text{ mode}:}$

i. $B(B^- \to K^- f_2) = 0.54 \times 10^{-6}$ is smaller than the experimental value $(1.3^{+0.4}_{-0.5}) \times 10^{-6}$. This decay mode is also likely to have contribution from the W-annihilation and W-exchange processes. Including the factorizable contribution of such diagrams, the decay amplitudes of $B^- \to K^- f_2$ get modified to (putting aside the scale factor $\frac{G_F}{\sqrt{2}} V_{ub} V_{us}^*$)

$$A(B^- \to K^- f_2) = \frac{1}{\sqrt{2}} a_1 f_K \cos\phi_T F^{B \to f_2}(m_K^2) +$$
$$\frac{1}{\sqrt{2}} a_1 f_B \cos\phi_T F^{f_2 \to K}(m_B^2). \quad (23)$$

As it is not possible to evaluate the form factor $F^{f_2 \to K}$ at $m_B^2$ even in the phenomenological models, it is treated as a free parameter. Taking $f_B = 0.176$ GeV, we find that the experimental branching ratio $B(B^- \to K^- f_2) = (1.3^{+0.4}_{-0.5}) \times 10^{-6}$ requires $F^{f_2 \to K}(m_B^2) = -0.083$ GeV. This value in turn enhances the branching ratio for $B^- \to K^- f_2$ through the W-annihilation contibution to $1.3 \times 10^{-6}$.

ii. Branching ratios of $B(B^- \to \eta K_2^-) = 1.2 \times 10^{-8}$ is small than the experimental value $(9.1 \pm 3.0) \times 10^{-6}$. Similar to $B^- \to K^- f_2$ decay, this decay mode is also likely to have contribution from the W-annihilation and W-exchange processes. Including the factorizable contribution of such diagrams, the decay amplitudes of $B \to \eta K_2$ get modified to (leaving aside the scale factor $\frac{G_F}{\sqrt{2}} V_{ub} V_{us}^*$)

$$A(B^- \to \eta K_2^-) = \frac{1}{\sqrt{2}} a_2 f_\eta \sin\phi_P F^{B \to K_2}(m_\eta^2) +$$
$$\frac{1}{\sqrt{2}} a_2 f_B \sin\phi_P F^{K_2 \to \eta}(m_B^2)$$
$$A(\overline{B}^0 \to \eta \overline{K}_2^0) = \frac{1}{\sqrt{2}} a_2 f_\eta \sin\phi_P F^{B \to K_2}(m_\eta^2) +$$
$$\frac{1}{\sqrt{2}} a_2 f_B \sin\phi_P F^{K_2 \to \eta}(m_B^2). \qquad (24)$$

For $f_B = 0.176$ GeV, we find that the experimental branching ratio $B(B^- \to \eta K_2^-) = (9.1 \pm 3.0) \times 10^{-6}$ requires $F^{K_2 \to \eta}(m_B^2) = -3.03$ GeV. This in turn enhances the branching ratio for $\overline{B}^0 \to \eta \overline{K}_2^0$ to $8.1 \times 10^{-6}$, which is consistent with the experimental value $(9.6 \pm 2.1) \times 10^{-6}$.

iii. Decays $B^- \to \pi^- \overline{K}_2^0 / \overline{K}^0 a_2^-$, $\overline{B}^0 \to \pi^+ K_2^- / \overline{K}^0 a_2^0 \ / \overline{K}^0 f_2 / \overline{K}^0 f_2'$ and $\overline{B}_s^0 \to K^+ K_2^- / K^0 \overline{K}_2^0 / \pi^+ a_2^- / \pi^0 a_2^0 / \pi^- a_2^+ \ / \eta a_2^0 / \overline{K}^0 K_2^0 / \eta' a_2^0$ are forbidden in the present analysis. Annihilation and FSIs may generate these decays.

## V. SUMMARY AND CONCLUSIONS

In this paper, we have studied hadronic weak decays of bottom mesons emitting pseudoscalar and tensor mesons. The matrix elements $\langle T(q_\mu)|J_\mu|0\rangle$ vanish due to the tracelessness of the polarization tensor $\in_{\mu\nu}$ of spin 2 meson and the auxiliary condition $q^\mu \in_{\mu\nu} = 0$. Therefore, either color-favored or color-suppressed diagrams contribute. Therefore, the analysis of these decays is free of constructive or destructive interference for color-favored and color-suppressed diagrams. We employ ISGW II model [3] to determine the $B \to T$ form factors appearing in the decay matrix element of weak currents involving $b \to c$ and $b \to u$ transitions. Consequently, we have obtained the decay amplitudes and calculated the branching ratios of $B \to PT$ decays in CKM-favored and CKM-suppressed modes. We make the following conclusions:

Decays involving $b \to c$ transition have larger branching ratios of the order of $10^{-4}$ to $10^{-8}$ and decays involving $b \to u$ transition have branching ratios of the order of $10^{-5}$ to $10^{-11}$. Dominant decay modes involving $b \to c$ transition are $B(B^- \to D_s^- D_2^0) = 6.8\times10^{-4}$, $B(B^- \to \pi^- D_2^0) = 6.7\times10^{-4}$, $B(\bar{B}^0 \to D_s^- D_2^+) = 6.4\times10^{-4}$, $B(\bar{B}^0 \to \pi^- D_2^+) = 6.1\times10^{-4}$, $B(B^- \to D^0 a_2^-) = 1.8\times10^{-4}$, $B(B^- \to \eta_c K_2^-) = 1.4\times10^{-4}$, $B(\bar{B}^0 \to \eta_c \bar{K}_2^0) = 1.3\times10^{-4}$, $B(\bar{B}_s^0 \to D_s^- D_{s2}^-) = 7.7\times10^{-4}$, $B(\bar{B}_s^0 \to \pi^- D_{s2}^+) = 7.1\times10^{-4}$, $B(\bar{B}_s^0 \to \eta_c f_2') = 1.3\times10^{-4}$ and $B(\bar{B}_s^0 \to D^0 K_2^0) = 1.1\times10^{-4}$. Experimentally, the branching ratios of only five decay modes are measured and upper limits are available for six other decays. We find that the calculated branching ratio $B(B^- \to \pi^- f_2) = 7.1\times10^{-6}$ is in good agreement with the experimental value $(8.2\pm2.5)\times10^{-6}$ whereas $B(B^- \to K^- f_2) = 5.4\times10^{-7}$ is smaller than the experimental value $(1.3^{+0.4}_{-0.5})\times10^{-6}$. $B$-decay requires contribution from W-annihilation diagram to bridge the gap between theoretical and experimental value. In contrast to the charm meson decays, the experimental data show constructive interference for $B$ meson decays involving both the color-favored and color-suppressed diagrams, giving $a_1 = 1.10\pm0.08$ and $a_2 = 0.20\pm0.02$. In the present analysis, the decay amplitude is proportional to only one QCD coefficient either $a_1$ (for color favored diagram) or $a_2$ (for color suppressed diagram), therefore our results remains unaffected from the interference pattern.

We also compare our results with branching ratios calculated in the other models [17,23,24]. The predicted branching ratios in KLO [17] shown in 3$^{rd}$ column of tables VI, VII, VIII(a) and VIII(b) are generally smaller as compared to the present branching ratios because of the difference in the form factors since different quark masses have been used in the two works. Branching ratios have also been calculated by Cheng [23]. His predictions $B(B^- \to \pi^- D_2^0) = 6.7\times10^{-4}$ and $B(\bar{B}^0 \to \pi^- D_2^+) = 6.1\times10^{-4}$ match well with the numerical branching ratios obtained in the present work. However, the other branching ratios $B(B^- \to D_s^- D_2^0) = 4.2\times10^{-4}$, $B(\bar{B}^0 \to D_s^- D_2^+) = 3.8\times10^{-4}$ and $B(\bar{B}_s^0 \to \pi^- D_{s2}^+)$ = 3.8×10$^{-4}$ are different from our results owing to the different values used for the decay constant $f_{D_s}$. MQ [24] have recently studied few charmless decays of $B \to PT$ mode. Some of the branching ratios are smaller than our numerical value of branching ratios while the others are large as compare to the present predictions, particularly, for $\eta, \eta'$ emitting decays. The disagreement with their predictions may be attributed due to the difference in the form factors obtained in covariant light-front approach (CLF) and inclusion of the non-factorizable contributions in their results. It may be noted that the form factors at small $q^2$ obtained in the CLF and ISGW II quark model agrees within 40% [3]. However, when $q^2$ increases $h(q^2)$, $b_+(q^2)$ and $b_-(q^2)$ increases more rapidly in the light front model than in the ISGW II model. Another important fact is that the behavior of the form factor $k$ in both models is different.

The Belle collaboration is currently searching for some $B \to PT$ modes and their preliminary results indicate that the branching ratios for these may not be very small compared to $B \to PP$ modes. We hope our predictions would be within the reach of the current experiments. Observation of these decays in the $B$ experiments such as Belle, Babar, BTeV, LHC and so on will be crucial in testing the ISGW II and other quark models as well as validity of the factorization scheme.

## Acknowledgement

One of the authors (N. S.) is thankful to the University Grant Commission, New Delhi, for the financial assistance.

**Table I. The parameter $\beta$ for *s*-wave and *p*-wave mesons in the ISGW II model**

| Quark content | $u\bar{d}$ | $u\bar{s}$ | $s\bar{s}$ | $c\bar{u}$ | $c\bar{s}$ | $u\bar{b}$ | $s\bar{b}$ |
|---|---|---|---|---|---|---|---|
| $\beta_S$ (GeV) | 0.41 | 0.44 | 0.53 | 0.45 | 0.56 | 0.43 | 0.54 |
| $\beta_P$ (GeV) | 0.28 | 0.30 | 0.33 | 0.33 | 0.38 | 0.35 | 0.41 |

**Table II. Form factors of $B \to T$ transition at $q^2 = t_m$ in the ISGW II quark model**

| Transition | $k$ | $b_+$ | $b_-$ |
|---|---|---|---|
| $B \to a_2$ | 0.432 | -0.013 | 0.015 |
| $B \to f_2$ | 0.425 | -0.014 | 0.014 |
| $B \to f_2'$ | 0.533 | -0.013 | 0.015 |
| $B \to K_2$ | 0.480 | -0.015 | 0.015 |
| $B \to D_2$ | 0.677 | -0.013 | 0.013 |
| $B_s \to f_2$ | 0.423 | -0.015 | 0.016 |
| $B_s \to f_2'$ | 0.572 | -0.016 | 0.017 |
| $B_s \to K_2$ | 0.492 | -0.013 | 0.015 |
| $B_s \to D_{s2}$ | 0.854 | -0.015 | 0.016 |

**Table III. Decay amplitudes of** $B \to PT$ **decays in CKM-favored mode involving** $b \to c$ **transition**

| Decay | Amplitude |
|---|---|
| $\Delta b = 1, \Delta C = 1, \Delta S = 0$ | $\times \dfrac{G_F}{\sqrt{2}} V_{cb} V_{ud}^*$ |
| $B^- \to \pi^- D_2^0$ | $a_1 f_\pi F^{B \to D_2}(m_\pi^2)$ |
| $B^- \to D^0 a_2^-$ | $a_2 f_D F^{B \to a_2}(m_D^2)$ |
| $\overline{B}^0 \to \pi^- D_2^+$ | $a_1 f_\pi F^{B \to D_2}(m_\pi^2)$ |
| $\overline{B}^0 \to D^0 a_2^0$ | $-\dfrac{1}{\sqrt{2}} a_2 f_D F^{B \to a_2}(m_D^2)$ |
| $\overline{B}^0 \to D^0 f_2$ | $\dfrac{1}{\sqrt{2}} a_2 f_D \cos\phi_T F^{B \to f_2}(m_D^2)$ |
| $\overline{B}^0 \to D^0 f_2'$ | $\dfrac{1}{\sqrt{2}} a_2 f_D \sin\phi_T F^{B \to f_2'}(m_D^2)$ |
| $\overline{B}_s^0 \to \pi^- D_{s2}^+$ | $a_1 f_\pi F^{B_s \to D_{s2}}(m_\pi^2)$ |
| $\overline{B}_s^0 \to D^0 K_2^0$ | $a_2 f_D F^{B_s \to K_2}(m_D^2)$ |
| $\Delta b = 1, \Delta C = 0, \Delta S = -1$ | $\times \dfrac{G_F}{\sqrt{2}} V_{cb} V_{cs}^*$ |
| $B^- \to D_s^- D_2^0$ | $a_1 f_{D_s} F^{B \to D_2}(m_{D_s}^2)$ |
| $B^- \to \eta_c K_2^-$ | $a_2 f_{\eta_c} F^{B \to K_2}(m_{\eta_c}^2)$ |
| $\overline{B}^0 \to D_s^- D_2^+$ | $a_1 f_{D_s} F^{B \to D_2}(m_{D_s}^2)$ |
| $\overline{B}^0 \to \eta_c K_2^0$ | $a_2 f_{\eta_c} F^{B \to K_2}(m_{\eta_c}^2)$ |
| $\overline{B}_s^0 \to D_s^- D_{s2}^+$ | $a_1 f_{D_s} F^{B_s \to D_{s2}}(m_{D_s}^2)$ |
| $\overline{B}_s^0 \to \eta_c f_2$ | $a_2 f_{\eta_c} \sin\phi_T F^{B_s \to f_2}(m_{\eta_c}^2)$ |
| $\overline{B}_s^0 \to \eta_c f_2'$ | $- a_2 f_{\eta_c} \cos\phi_T F^{B_s \to f_2'}(m_{\eta_c}^2)$ |

**Table IV. Decay amplitudes of** $B \to PT$ **decays in CKM-suppressed mode involving** $b \to c$ **transition**

| Decay | Amplitude |
|---|---|
| $\Delta b = 1, \Delta C = 1, \Delta S = -1$ | $\times \dfrac{G_F}{\sqrt{2}} V_{cb} V_{us}^*$ |
| $B^- \to K^- D_2^0$ | $a_1 f_K F^{B \to D_2}(m_K^2)$ |
| $B^- \to D^0 K_2^-$ | $a_2 f_D F^{B \to K_2}(m_D^2)$ |
| $\overline{B}^0 \to K^- D_2^+$ | $a_1 f_K F^{B \to D_2}(m_K^2)$ |
| $\overline{B}^0 \to D^0 \overline{K}_2^0$ | $a_2 f_D F^{B \to K_2}(m_D^2)$ |
| $\overline{B}_s^0 \to K^- D_{s2}^+$ | $a_1 f_K F^{B_s \to D_{s2}}(m_K^2)$ |
| $\overline{B}_s^0 \to D^0 f_2$ | $a_2 f_D \sin\phi_T F^{B_s \to f_2}(m_D^2)$ |
| $\overline{B}_s^0 \to D^0 f_2'$ | $-a_2 f_D \cos\phi_T F^{B_s \to f_2'}(m_D^2)$ |
| $\Delta b = 1, \Delta C = 0, \Delta S = 0$ | $\times \dfrac{G_F}{\sqrt{2}} V_{cb} V_{cd}^*$ |
| $B^- \to D^- D_2^0$ | $a_1 f_D F^{B \to D_2}(m_D^2)$ |
| $B^- \to \eta_c a_2^-$ | $a_2 f_{\eta_c} F^{B \to a_2}(m_{\eta_c}^2)$ |
| $\overline{B}^0 \to D^- D_2^+$ | $a_1 f_D F^{B \to D_2}(m_D^2)$ |
| $\overline{B}^0 \to \eta_c a_2^0$ | $-\dfrac{1}{\sqrt{2}} a_2 f_{\eta_c} F^{B \to a_2}(m_{\eta_c}^2)$ |
| $\overline{B}^0 \to \eta_c f_2$ | $\dfrac{1}{\sqrt{2}} a_2 f_{\eta_c} \cos\phi_T F^{B \to f_2}(m_{\eta_c}^2)$ |
| $\overline{B}^0 \to \eta_c f_2'$ | $\dfrac{1}{\sqrt{2}} a_2 f_{\eta_c} \sin\phi_T F^{B \to f_2'}(m_{\eta_c}^2)$ |
| $\overline{B}_s^0 \to D^- D_{s2}^+$ | $a_1 f_D F^{B_s \to D_{s2}}(m_D^2)$ |
| $\overline{B}_s^0 \to \eta_c K_2^0$ | $a_2 f_{\eta_c} F^{B_s \to K_2}(m_{\eta_c}^2)$ |

**Table V (a). Decay amplitudes of** $B \to PT$ **decays involving** $b \to u$ **transition**

| Decay | Amplitude |
|---|---|
| $\Delta b = 1, \Delta C = -1, \Delta S = -1$ | $\times \dfrac{G_F}{\sqrt{2}} V_{ub} V_{cs}^*$ |
| $B^- \to \overline{D}^0 K_2^-$ | $a_2 f_D F^{B \to K_2}(m_D^2)$ |
| $B^- \to D_s^- a_2^0$ | $\dfrac{1}{\sqrt{2}} a_1 f_{D_s} F^{B \to a_2}(m_{D_s}^2)$ |
| $B^- \to D_s^- f_2$ | $\dfrac{1}{\sqrt{2}} a_1 f_{D_s} \cos\phi_T F^{B \to f_2}(m_{D_s}^2)$ |
| $B^- \to D_s^- f_2'$ | $\dfrac{1}{\sqrt{2}} a_1 f_{D_s} \sin\phi_T F^{B \to f_2'}(m_{D_s}^2)$ |
| $\overline{B}^0 \to \overline{D}^0 \overline{K}_2^0$ | $a_2 f_D F^{B \to K_2}(m_D^2)$ |
| $\overline{B}^0 \to D_s^- a_2^+$ | $a_1 f_{D_s} F^{B \to a_2}(m_{D_s}^2)$ |
| $\overline{B}_s^0 \to \overline{D}^0 f_2$ | $a_2 f_D \sin\phi_T F^{B_s \to f_2}(m_D^2)$ |
| $\overline{B}_s^0 \to \overline{D}^0 f_2'$ | $-a_2 f_D \cos\phi_T F^{B_s \to f_2}(m_D^2)$ |
| $\overline{B}_s^0 \to D_s^- K_2^+$ | $a_1 f_{D_s} F^{B_s \to K_2}(m_{D_s}^2)$ |
| $\Delta b = 1, \Delta C = 0, \Delta S = 0$ | $\times \dfrac{G_F}{\sqrt{2}} V_{ub} V_{ud}^*$ |
| $B^- \to \pi^0 a_2^-$ | $\dfrac{1}{\sqrt{2}} a_2 f_\pi F^{B \to a_2}(m_\pi^2)$ |
| $B^- \to \eta a_2^-$ | $\dfrac{1}{\sqrt{2}} a_2 f_\eta \sin\phi_P F^{B \to a_2}(m_\eta^2)$ |
| $B^- \to \eta' a_2^-$ | $\dfrac{1}{\sqrt{2}} a_2 f_{\eta'} \cos\phi_P F^{B \to a_2}(m_{\eta'}^2)$ |
| $B^- \to \pi^- a_2^0$ | $\dfrac{1}{\sqrt{2}} a_1 f_\pi F^{B \to a_2}(m_\pi^2)$ |
| $B^- \to \pi^- f_2$ | $\dfrac{1}{\sqrt{2}} a_1 f_\pi \cos\phi_T F^{B \to f_2}(m_\pi^2)$ |
| $B^- \to \pi^- f_2'$ | $\dfrac{1}{\sqrt{2}} a_1 f_\pi \sin\phi_T F^{B \to f_2'}(m_\pi^2)$ |
| $\overline{B}^0 \to \pi^- a_2^+$ | $a_1 f_\pi F^{B \to a_2}(m_\pi^2)$ |
| $\overline{B}^0 \to \pi^0 a_2^0$ | $-\dfrac{1}{2} a_2 f_\pi F^{B \to a_2}(m_\pi^2)$ |
| $\overline{B}^0 \to \pi^0 f_2$ | $\dfrac{1}{2} a_2 f_\pi \cos\phi_T F^{B\to f_2}(m_\pi^2)$ |
| $\overline{B}^0 \to \pi^0 f_2'$ | $\dfrac{1}{2} a_2 f_\pi \sin\phi_T F^{B\to f_2'}(m_\pi^2)$ |
| $\overline{B}^0 \to \eta a_2^0$ | $-\dfrac{1}{2} a_2 f_\eta \sin\phi_P F^{B\to a_2}(m_\eta^2)$ |
| $\overline{B}^0 \to \eta f_2$ | $\dfrac{1}{2} a_2 f_\eta \sin\phi_P \cos\phi_T F^{B\to f_2}(m_\eta^2)$ |
| $\overline{B}^0 \to \eta f_2'$ | $\dfrac{1}{2} a_2 f_\eta \sin\phi_P \sin\phi_T F^{B\to f_2'}(m_\eta^2)$ |
| $\overline{B}^0 \to \eta' a_2^0$ | $-\dfrac{1}{2} a_2 f_{\eta'} \cos\phi_P F^{B\to a_2}(m_{\eta'}^2)$ |
| $\overline{B}^0 \to \eta' f_2$ | $\dfrac{1}{2} a_2 f_{\eta'} \cos\phi_P \cos\phi_T F^{B\to f_2}(m_{\eta'}^2)$ |
| $\overline{B}^0 \to \eta' f_2'$ | $\dfrac{1}{2} a_2 f_{\eta'} \cos\phi_P \sin\phi_T F^{B\to f_2'}(m_{\eta'}^2)$ |
| $\overline{B}_s^0 \to \pi^0 K_2^0$ | $\dfrac{1}{\sqrt{2}} a_2 f_\pi F^{B_s\to K_2}(m_\pi^2)$ |
| $\overline{B}_s^0 \to \pi^- K_2^+$ | $a_1 f_\pi F^{B_s\to K_2}(m_\pi^2)$ |
| $\overline{B}_s^0 \to \eta K_2^0$ | $\dfrac{1}{\sqrt{2}} a_2 f_\eta \sin\phi_P F^{B_s\to K_2}(m_\eta^2)$ |
| $\overline{B}_s^0 \to \eta' K_2^0$ | $\dfrac{1}{\sqrt{2}} a_2 f_{\eta'} \cos\phi_P F^{B_s\to K_2}(m_{\eta'}^2)$ |

**Table V (b). Decay amplitudes of $B \to PT$ decays involving $b \to u$ transition**

| Decay | Amplitude |
|---|---|
| $\Delta b = 1, \Delta C = 0, \Delta S = -1$ | $\times \dfrac{G_F}{\sqrt{2}} V_{ub} V_{us}^*$ |
| $B^- \to K^- a_2^0$ | $\dfrac{1}{\sqrt{2}} a_1 f_K F^{B \to a_2}(m_K^2)$ |
| $B^- \to K^- f_2$ | $\dfrac{1}{\sqrt{2}} a_1 f_K \cos\phi_T F^{B \to f_2}(m_K^2)$ |
| $B^- \to K^- f_2'$ | $\dfrac{1}{\sqrt{2}} a_1 f_K \sin\phi_T F^{B \to f_2'}(m_K^2)$ |
| $B^- \to \pi^0 K_2^-$ | $\dfrac{1}{\sqrt{2}} a_2 f_\pi F^{B \to K_2}(m_\pi^2)$ |
| $B^- \to \eta K_2^-$ | $\dfrac{1}{\sqrt{2}} a_2 f_\eta \sin\phi_P F^{B \to K_2}(m_\eta^2)$ |
| $B^- \to \eta' K_2^-$ | $\dfrac{1}{\sqrt{2}} a_2 f_{\eta'} \cos\phi_P F^{B \to K_2}(m_{\eta'}^2)$ |
| $\bar{B}^0 \to K^- a_2^+$ | $a_1 f_K F^{B \to a_2}(m_K^2)$ |
| $\bar{B}^0 \to \pi^0 \bar{K}_2^0$ | $\dfrac{1}{\sqrt{2}} a_2 f_\pi F^{B \to K_2}(m_\pi^2)$ |
| $\bar{B}^0 \to \eta \bar{K}_2^0$ | $\dfrac{1}{\sqrt{2}} a_2 f_\eta \sin\phi_P F^{B \to K_2}(m_\eta^2)$ |
| $\bar{B}^0 \to \eta' \bar{K}_2^0$ | $\dfrac{1}{\sqrt{2}} a_2 f_{\eta'} \cos\phi_P F^{B \to K_2}(m_{\eta'}^2)$ |
| $\bar{B}_s^0 \to \pi^0 f_2$ | $\dfrac{1}{\sqrt{2}} a_2 f_\pi \sin\phi_T F^{B_s \to f_2}(m_\pi^2)$ |
| $\bar{B}_s^0 \to \pi^0 f_2'$ | $-\dfrac{1}{\sqrt{2}} a_2 f_\pi \cos\phi_T F^{B_s \to f_2'}(m_\pi^2)$ |
| $\bar{B}_s^0 \to \eta f_2$ | $\dfrac{1}{\sqrt{2}} a_2 f_\eta \sin\phi_P \sin\phi_T F^{B_s \to f_2}(m_\eta^2)$ |
| $\bar{B}_s^0 \to K^- K_2^+$ | $a_1 f_K F^{B_s \to K_2}(m_K^2)$ |
| $\bar{B}_s^0 \to \eta f_2'$ | $-\dfrac{1}{\sqrt{2}} a_2 f_\eta \sin\varphi_P \cos\varphi_T F^{B_s \to f_2'}(m_\eta^2)$ |
| $\bar{B}_s^0 \to \eta' f_2$ | $\dfrac{1}{\sqrt{2}} a_2 f_{\eta'} \cos\phi_P \sin\phi_T F^{B_s \to f_2}(m_{\eta'}^2)$ |
| $\bar{B}_s^0 \to \eta' f_2'$ | $-\dfrac{1}{\sqrt{2}} a_2 f_{\eta'} \cos\phi_P \cos\phi_T F^{B_s \to f_2'}(m_{\eta'}^2)$ |

| $\Delta b = 1, \Delta C = -1, \Delta S = 0$ | $\times \dfrac{G_F}{\sqrt{2}} V_{ub} V_{cd}^*$ |
| --- | --- |
| $B^- \to D^- a_2^0$ | $\dfrac{1}{\sqrt{2}} a_1 f_D F^{B \to a_2}(m_D^2)$ |
| $B^- \to D^- f_2$ | $\dfrac{1}{\sqrt{2}} a_1 f_D \cos\phi_T F^{B \to f_2}(m_D^2)$ |
| $B^- \to D^- f_2'$ | $\dfrac{1}{\sqrt{2}} a_1 f_D \sin\phi_T F^{B \to f_2'}(m_D^2)$ |
| $B^- \to \overline{D}^0 a_2^-$ | $a_2 f_D F^{B \to a_2}(m_D^2)$ |
| $\overline{B}^0 \to \overline{D}^0 a_2^0$ | $-\dfrac{1}{\sqrt{2}} a_2 f_D F^{B \to a_2}(m_D^2)$ |
| $\overline{B}^0 \to \overline{D}^0 f_2$ | $\dfrac{1}{\sqrt{2}} a_2 f_D \cos\phi_T F^{B \to f_2}(m_D^2)$ |
| $\overline{B}^0 \to \overline{D}^0 f_2'$ | $\dfrac{1}{\sqrt{2}} a_2 f_D \sin\phi_T F^{B \to f_2'}(m_D^2)$ |
| $\overline{B}^0 \to D^- a_2^+$ | $a_1 f_D F^{B \to a_2}(m_D^2)$ |
| $\overline{B}_s^0 \to D^- K_2^+$ | $a_1 f_D F^{B_s \to K_2}(m_D^2)$ |
| $\overline{B}_s^0 \to \overline{D}^0 K_2^0$ | $a_2 f_D F^{B_s \to K_2}(m_D^2)$ |

**Table VI. Branching ratios of** $B \to PT$ **decays in CKM-favored mode involving** $b \to c$ **transition**

| Decay | Branching ratios | |
|---|---|---|
| | This Work | KLO |
| $\Delta b = 1, \Delta C = 1, \Delta S = 0$ | | |
| $B^- \to \pi^- D_2^0$ | $6.7 \times 10^{-4}$ | $3.5 \times 10^{-4}$ |
| $B^- \to D^0 a_2^-$ | $1.8 \times 10^{-4}$ | $1.0 \times 10^{-4}$ |
| $\bar{B}^0 \to \pi^- D_2^+$ | $6.1 \times 10^{-4}$ | $3.3 \times 10^{-4}$ |
| $\bar{B}^0 \to D^0 a_2^0$ | $8.2 \times 10^{-5}$ | $4.8 \times 10^{-4}$ |
| $\bar{B}^0 \to D^0 f_2$ | $8.8 \times 10^{-5}$ | $5.3 \times 10^{-5}$ |
| $\bar{B}^0 \to D^0 f_2'$ | $1.7 \times 10^{-6}$ | $0.62 \times 10^{-6}$ |
| $\bar{B}_s^0 \to \pi^- D_{s2}^+$ | $7.1 \times 10^{-4}$ | - |
| $\bar{B}_s^0 \to D^0 K_2^0$ | $1.1 \times 10^{-4}$ | - |
| $\Delta b = 1, \Delta C = 0, \Delta S = -1$ | | |
| $B^- \to D_s^- D_2^0$ | $6.8 \times 10^{-4}$ | $4.9 \times 10^{-4}$ |
| $B^- \to \eta_c K_2^-$ | $1.4 \times 10^{-4}$ | $1.1 \times 10^{-4}$ |
| $\bar{B}^0 \to D_s^- D_2^+$ | $6.4 \times 10^{-4}$ | $4.6 \times 10^{-4}$ |
| $\bar{B}^0 \to \eta_c \bar{K}_2^0$ | $1.3 \times 10^{-4}$ | $9.6 \times 10^{-5}$ |
| $\bar{B}_s^0 \to D_s^- D_{s2}^-$ | $7.7 \times 10^{-4}$ | - |
| $\bar{B}_s^0 \to \eta_c f_2$ | $2.7 \times 10^{-6}$ | - |
| $\bar{B}_s^0 \to \eta_c f_2'$ | $1.3 \times 10^{-4}$ | - |

**Table VII. Branching ratios of** $B \to PT$ **decays in CKM-suppressed mode involving** $b \to c$ **transition**

| Decay | Branching ratios | |
|---|---|---|
| | **This Work** | **KLO** |
| $\Delta b = 1, \Delta C = 1, \Delta S = -1$ | | |
| $B^- \to K^- D_2^0$ | $4.8 \times 10^{-5}$ | $2.5 \times 10^{-5}$ |
| $B^- \to D^0 K_2^-$ | $8.7 \times 10^{-6}$ | $7.3 \times 10^{-6}$ |
| $\overline{B}^0 \to K^- D_2^+$ | $4.5 \times 10^{-5}$ | $2.4 \times 10^{-5}$ |
| $\overline{B}^0 \to D^0 \overline{K}_2^0$ | $8.1 \times 10^{-6}$ | $6.8 \times 10^{-6}$ |
| $\overline{B}_s^0 \to K^- D_{s2}^+$ | $5.2 \times 10^{-5}$ | - |
| $\overline{B}_s^0 \to D^0 f_2$ | $9.9 \times 10^{-8}$ | - |
| $\overline{B}_s^0 \to D^0 f_2'$ | $6.7 \times 10^{-6}$ | - |
| $\Delta b = 1, \Delta C = 0, \Delta S = 0$ | | |
| $B^- \to D^- D_2^0$ | $2.5 \times 10^{-5}$ | $2.2 \times 10^{-5}$ |
| $B^- \to \eta_c a_2^-$ | $9.2 \times 10^{-6}$ | $4.9 \times 10^{-6}$ |
| $\overline{B}^0 \to D^- D_2^+$ | $2.4 \times 10^{-5}$ | $2.1 \times 10^{-5}$ |
| $\overline{B}^0 \to \eta_c a_2^0$ | $4.3 \times 10^{-6}$ | $2.3 \times 10^{-6}$ |
| $\overline{B}^0 \to \eta_c f_2$ | $4.8 \times 10^{-6}$ | $2.7 \times 10^{-6}$ |
| $\overline{B}^0 \to \eta_c f_2'$ | $6.7 \times 10^{-8}$ | $0.02 \times 10^{-6}$ |
| $\overline{B}_s^0 \to D^- D_{s2}^+$ | $2.9 \times 10^{-5}$ | - |
| $\overline{B}_s^0 \to \eta_c K_2^0$ | $6.9 \times 10^{-6}$ | - |

**Table VIII (a). Branching ratios of $B \to PT$ decays involving $b \to u$ transition**

| Decays | Branching ratios | | |
|---|---|---|---|
| | This Work | KLO | MQ |
| $\Delta b = 1, \Delta C = -1, \Delta S = -1$ | | | |
| $B^- \to \overline{D}^0 K_2^-$ | $1.3 \times 10^{-6}$ | $1.2 \times 10^{-6}$ | - |
| $B^- \to D_s^- a_2^0$ | $2.0 \times 10^{-5}$ | $9.4 \times 10^{-6}$ | - |
| $B^- \to D_s^- f_2$ | $2.2 \times 10^{-5}$ | $1. \times 10^{-5}$ | - |
| $B^- \to D_s^- f_2'$ | $4.3 \times 10^{-7}$ | $0.12 \times 10^{-6}$ | - |
| $\overline{B}^0 \to \overline{D}^0 \overline{K}_2^0$ | $1.2 \times 10^{-6}$ | $1.1 \times 10^{-6}$ | - |
| $\overline{B}^0 \to D_s^- a_2^+$ | $3.8 \times 10^{-5}$ | $1.8 \times 10^{-5}$ | - |
| $\overline{B}_s^0 \to D_s^- K_2^+$ | $2.6 \times 10^{-5}$ | - | - |
| $\overline{B}_s^0 \to \overline{D}^0 f_2$ | $1.5 \times 10^{-8}$ | - | - |
| $\overline{B}_s^0 \to \overline{D}^0 f_2'$ | $1.0 \times 10^{-6}$ | - | - |
| $\Delta b = 1, \Delta C = 0, \Delta S = 0$ | | | |
| $B^- \to \pi^- a_2^0$ | $6.7 \times 10^{-6}$ | $2.6 \times 10^{-6}$ | $4.38 \times 10^{-6}$ |
| $B^- \to \pi^- f_2$ | $7.1 \times 10^{-6}$ | - | - |
| $B^- \to \pi^- f_2'$ | $1.5 \times 10^{-7}$ | - | - |
| $B^- \to \pi^0 a_2^-$ | $0.38 \times 10^{-6}$ | $0.001 \times 10^{-6}$ | $0.015 \times 10^{-6}$ |
| $B^- \to \eta a_2^-$ | $0.23 \times 10^{-6}$ | $0.29 \times 10^{-6}$ | $45.8 \times 10^{-6}$ |
| $B^- \to \eta' a_2^-$ | $0.13 \times 10^{-6}$ | $1.31 \times 10^{-6}$ | $71.3 \times 10^{-6}$ |
| $\overline{B}^0 \to \pi^- a_2^+$ | $13.0 \times 10^{-6}$ | $4.88 \times 10^{-6}$ | $8.19 \times 10^{-6}$ |
| $\overline{B}^0 \to \pi^0 a_2^0$ | $0.18 \times 10^{-6}$ | $0.0003 \times 10^{-6}$ | $0.007 \times 10^{-6}$ |
| $\overline{B}^0 \to \pi^0 f_2$ | $1.9 \times 10^{-7}$ | - | - |
| $\overline{B}^0 \to \pi^0 f_2'$ | $3.9 \times 10^{-9}$ | - | - |
| $\overline{B}^0 \to \eta a_2^0$ | $0.11 \times 10^{-6}$ | $0.14 \times 10^{-6}$ | $25.2 \times 10^{-6}$ |
| $\overline{B}^0 \to \eta f_2$ | $1.1 \times 10^{-7}$ | - | - |
| $\overline{B}^0 \to \eta f_2'$ | $2.4 \times 10^{-9}$ | - | - |
| $\overline{B}^0 \to \eta' a_2^0$ | $0.06 \times 10^{-6}$ | $0.62 \times 10^{-6}$ | $43.3 \times 10^{-6}$ |
| $\overline{B}^0 \to \eta' f_2$ | $6.3 \times 10^{-8}$ | - | - |
| $\overline{B}^0 \to \eta' f_2'$ | $1.3 \times 10^{-9}$ | - | - |
| $\overline{B}_s^0 \to \pi^- K_2^+$ | $7.8 \times 10^{-6}$ | - | - |
| $\overline{B}_s^0 \to \pi^0 K_2^0$ | $2.2 \times 10^{-7}$ | - | - |
| $\overline{B}_s^0 \to \eta K_2^0$ | $1.3 \times 10^{-7}$ | - | - |
| $\overline{B}_s^0 \to \eta' K_2^0$ | $7.5 \times 10^{-8}$ | - | - |

**Table VIII (b). Branching ratios of $B \to PT$ decays involving $b \to u$ transition**

| Decays | Branching ratios | | |
|---|---|---|---|
| | This Work | KLO | MQ |
| $\Delta b = 1, \Delta C = 0, \Delta S = -1$ | | | |
| $B^- \to K^- a_2^0$ | $0.51 \times 10^{-6}$ | $0.31 \times 10^{-6}$ | $0.39 \times 10^{-6}$ |
| $B^- \to K^- f_2$ | $5.4 \times 10^{-7}$ | - | - |
| $B^- \to K^- f_2'$ | $1.5 \times 10^{-8}$ | - | - |
| $B^- \to \pi^0 K_2^-$ | $0.02 \times 10^{-6}$ | $0.09 \times 10^{-6}$ | $0.15 \times 10^{-6}$ |
| $B^- \to \eta K_2^-$ | $0.01 \times 10^{-6}$ | $0.03 \times 10^{-6}$ | $1.19 \times 10^{-6}$ |
| $B^- \to \eta' K_2^-$ | $0.007 \times 10^{-6}$ | $1.40 \times 10^{-6}$ | $2.70 \times 10^{-6}$ |
| $\overline{B}^0 \to K^- a_2^+$ | $0.95 \times 10^{-6}$ | $0.58 \times 10^{-6}$ | $0.73 \times 10^{-6}$ |
| $\overline{B}^0 \to \pi^0 \overline{K}_2^0$ | $0.02 \times 10^{-6}$ | $0.08 \times 10^{-6}$ | $0.13 \times 10^{-6}$ |
| $\overline{B}^0 \to \eta \overline{K}_2^0$ | $0.01 \times 10^{-6}$ | $0.03 \times 10^{-6}$ | $1.09 \times 10^{-6}$ |
| $\overline{B}^0 \to \eta' \overline{K}_2^0$ | $0.006 \times 10^{-6}$ | $1.3 \times 10^{-6}$ | $2.46 \times 10^{-6}$ |
| $\overline{B}_s^0 \to K^- K_2^+$ | $5.9 \times 10^{-7}$ | - | |
| $\overline{B}_s^0 \to \pi^0 f_2$ | $1.9 \times 10^{-10}$ | - | |
| $\overline{B}_s^0 \to \pi^0 f_2'$ | $1.4 \times 10^{-8}$ | - | |
| $\overline{B}_s^0 \to \eta f_2$ | $1.1 \times 10^{-10}$ | - | |
| $\overline{B}_s^0 \to \eta f_2'$ | $8.3 \times 10^{-9}$ | - | |
| $\overline{B}_s^0 \to \eta' f_2$ | $6.5 \times 10^{-11}$ | - | |
| $\overline{B}_s^0 \to \eta' f_2'$ | $4.7 \times 10^{-9}$ | - | |
| $\Delta b = 1, \Delta C = -1, \Delta S = 0$ | | | |
| $B^- \to D^- a_2^0$ | $6.5 \times 10^{-7}$ | - | |
| $B^- \to D^- f_2$ | $6.9 \times 10^{-7}$ | - | |
| $B^- \to D^- f_2'$ | $1.4 \times 10^{-7}$ | - | |
| $B^- \to \overline{D}^0 a_2^-$ | $7.3 \times 10^{-8}$ | - | |
| $\overline{B}^0 \to D^- a_2^+$ | $1.2 \times 10^{-6}$ | - | |
| $\overline{B}^0 \to \overline{D}^0 a_2^0$ | $3.4 \times 10^{-8}$ | - | |
| $\overline{B}^0 \to \overline{D}^0 f_2$ | $3.6 \times 10^{-8}$ | - | |
| $\overline{B}^0 \to \overline{D}^0 f_2'$ | $7.1 \times 10^{-10}$ | - | |
| $\overline{B}_s^0 \to D^- K_2^+$ | $8.3 \times 10^{-7}$ | - | |
| $\overline{B}_s^0 \to \overline{D}^0 K_2^0$ | $4.6 \times 10^{-8}$ | - | |